# Does VO$_2$ Host a Transient Monoclinic Metallic Phase?

Luciana Vidas,[1] Daniel Schick,[1,2,*] Elías Martínez,[1] Daniel Perez-Salinas,[1] Alberto Ramos-Álvarez,[1] Simon Cichy,[1] Sergi Batlle-Porro,[1] Allan S. Johnson,[1] Kent A. Hallman,[3,†] Richard F. Haglund, Jr.,[3] and Simon Wall[1,4,‡]

[1]*ICFO-Institut de Ciències Fotòniques, The Barcelona Institute of Science and Technology,*
*08860 Castelldefels, Barcelona, Spain*
[2]*Helmholtz-Zentrum Berlin für Materialien und Energie GmbH,*
*Albert-Einstein-Strasse 15, 12489 Berlin, Germany*
[3]*Department of Physics and Astronomy, Vanderbilt University,*
*Nashville, Tennessee 37235-1807, United States*
[4]*Department of Physics and Astronomy, Aarhus University, Ny Munkegade 120, 8000 Aarhus C, Denmark*

Ultrafast phase transitions induced by femtosecond light pulses present a new opportunity for manipulating the properties of materials. Understanding how these transient states are different from, or similar to, their thermal counterparts is key to determining how materials can exhibit properties that are not found in equilibrium. In this paper, we reexamine the case of the light-induced insulator-metal phase transition in the prototypical, strongly correlated material VO$_2$, for which a nonthermal Mott-Hubbard transition has been claimed. Here, we show that heat, even on the ultrafast timescale, plays a key role in the phase transition. When heating is properly accounted for, we find a single phase-transition threshold corresponding to the thermodynamic structural insulator-metal phase transition, and we find no evidence of a hidden transient Mott-Hubbard nonthermal phase. The interplay between the initial thermal state and the ultrafast transition may have implications for other transient states of matter.

## I. INTRODUCTION

Photoinduced phase transitions open up new ways to manipulate and probe the properties of materials, and ultrashort optical pulses can drive materials rapidly across phase boundaries in multiple ways. In most scenarios, the laser pulse rapidly excites the electrons, and a thermodynamic temperature commensurate with the absorbed energy density is established after the hot electrons reach equilibrium with the cooler ions on some timescale. If the induced temperature jump is sufficient to bring the system above a thermodynamic critical temperature, a phase transition occurs. The resulting state may be metastable if the temperature jump induces melting or disorders the solid, and the subsequent cooling is rapid enough to quench the disordered state at the initial temperature. These states can be stable for days or longer and can be characterized by a range of techniques after the transitions occur [1–4]. Alternatively, the high-symmetry state may only persist while the material is hot, and as the material cools, the original phase is restored. In this case, all the properties of the transient phase must be measured during the dynamical transition. In addition to these rapid but quasithermal transitions, many transient phases have been claimed that do not have thermodynamic analogues and result from the nonequilibrium process of the excitation mechanism. Often, these phases exhibit highly desirable and exotic phenomena, the most spectacular being light-induced superconductivity [5]. Understanding how and why such nonthermal phases form and how they connect to thermodynamic phases of the material is an outstanding and unsolved challenge in condensed matter physics.

Vanadium dioxide (VO$_2$) is the ideal material in which to study these effects. It is considered a Mott insulator, with an electronic insulator-to-metal transition accompanied by a monoclinic-to-rutile (M$_1 \rightarrow$ R) symmetry structural change. The phase transition in strain-free single crystals occurs at $T_c = 65\,°C$ [6]. The ultrafast photoinduced phase transition only occurs when the pump pulse exceeds a critical fluence threshold, and as one of the first and most studied materials, the initial concepts coined from studying this material have shaped the vocabulary used to understand a broad range of photoinduced phase transitions. However, there is still significant debate over the nature of the ultrafast phase transition in $VO_2$, in particular, whether the transient phase generated out of the insulating state is the same as the rutile metal, or if a new metallic state can be formed without a structural phase transformation. Central to understanding this problem is determining if the fluence threshold corresponds to a critical energy density or a critical number of excited electrons and how this relates to the thermal energy for the phase transition.

Recent experiments have suggested that a transient monoclinic metallic phase forms in $VO_2$ after photoexcitation. The threshold fluence required to make the system metallic, as deduced from mid-IR and THz probes, was observed to be lower than the fluence required to drive the structural transition as determined by electron diffraction. Thus, a new phase was proposed, which has metallic electronic properties but retains the $M_1$ crystal structure [7,8]. Such findings are consistent with density functional theory calculations that show how the band gap can close through screening changes induced by excited charges alone, without the need for the lattice to change [9]. While other theoretical work has shown that a sufficient number of photoexcited holes can destabilize the lattice and drive the structural transition [10,11], it is not unreasonable to conclude that photoexcited carriers may, in fact, initiate a monoclinic metallic phase at a lower excitation threshold than that required to destabilize the structural distortion.

However, identifying the monoclinic phase requires comparing results from different experimental setups, a process that can introduce systematic errors. This can occur even in the comparatively simple case of equilibrium measurements. For instance, comparing near-field imaging and x-ray diffraction methods gave rise to the claim that the electronic and structural phase transitions in the related vanadium compound $V_2O_3$ are decoupled in temperature [12], while later measurements showed this to be an artifact arising from comparing techniques [13]. The equilibrium transition is similarly controversial for $VO_2$, where approaches that combine different techniques have suggested that the equilibrium transitions can be decoupled [14], while techniques that can probe both aspects of the phase transition simultaneously show they are not [15]. Similar issues may then also affect ultrafast measurements.

In this paper, we reexamine the fluence dependence of the phase transition and look for systematic errors that may arise from comparing different techniques. We show that the freestanding samples in which the monoclinic metallic phase has been reported are extremely sensitive to heat accumulation, which results in inaccurate measurements of the threshold fluence. Furthermore, we show that, when using techniques such as x-ray absorption spectroscopy (XAS) and broadband optical absorption, which are sensitive to both electronic and structural changes in a single technique, only one fluence threshold is found for both degrees of freedom. Our results show that there is a single fluence threshold at which the monoclinic insulator is transformed into the rutile metallic phase.

The lack of a monoclinic metallic phase suggests that electronic correlations alone cannot be melted in $VO_2$. Even though the phase transition occurs on the ultrafast timescale, both the initial thermal energy in the system and the absorbed laser energy are determinative of the phase transition. Our results show that the thermal energy in the lattice, which, to date, has been neglected in models of $VO_2$, has a much stronger influence on the ultrafast phase transition than any other degree of freedom.

## II. HEATING AS A SOURCE FOR SYSTEMATIC ERRORS

Although a dependence on the threshold fluence for $VO_2$ with the base temperature has been known for some time [16,17], the implications of these results have yet to be fully examined. In particular, in order to measure an accurate fluence dependence for the phase transition, it is crucial that the sample recovers to the same initial temperature after photoexcitation independent of the excitation fluence. In the following, we show that this does not occur in freestanding samples, which are typically used for electron diffraction or x-ray absorption experiments, even at Hz repetition rates. The lack of thermal recovery results in an unreliable fluence dependence which can explain the apparent differences in the obtained threshold values when measured with different techniques.

The cooling rate determines the recovery time of a sample and is given by the thermal gradient. For thick samples, or thin samples on a substrate, the dominant thermal gradient points in the direction of the surface normal, and it is set by the pump penetration depth or film thickness. This typically leads to steep temperature gradients over a few hundred nanometers and results in rapid cooling of the volume of interest. However, in freestanding films, the dominant gradient is in the plane of the material. This case is now set by the pump spot size, which is typically tens to hundreds of microns, resulting in cooling dynamics many orders of magnitude slower. For example, in a two-dimensional system, the time required for the central region of the sample to cool down to a temperature $T_f$ after being excited to an initial temperature $T_0$ by a pump beam with waist $w$ is

$$t_f = \frac{w^2}{\alpha}\left(\frac{T_0}{T_f} - 1\right), \quad (1)$$

where $\alpha$ is the thermal diffusivity of the material, which is $2 \times 10^{-6}$ m$^2$ s$^{-1}$ for $VO_2$ [18]. If the pump beam has a waist of 30 μm (≈100 μm FWHM), the time taken for the temperature to fall by 90% of the initial jump is 4 ms. This amount is significantly longer than the approximately 10-ns timescales required for bulk samples or thin films on a substrate, and it has been confirmed by heat-flow simulations in real experimental geometries [18]. The repetition rate of pump-probe experiments should then be much less than 250 Hz in order to avoid heat accumulation. While a smaller pump focus would increase the cooling rate, many experimental probes require even larger sizes than those considered here. For instance, THz conductivity measurements require pump spot sizes at least a factor of 2 larger than the longest probe wavelength. As 1 THz has a 300-μm wavelength, cooling rates can be 50 times longer than described above and would require sub-Hz repetition rates. On these timescales, avoiding laser-induced heat accumulation is thus extremely challenging. In the next section, we quantify this heating effect and show how it can give rise to errors in the analysis of the phase transition.

### A. Measuring heat dynamics in thin VO$_2$

We study a 70-nm-thick polycrystalline VO$_2$ film deposited by pulsed laser deposition onto a 150-nm-thick freestanding Si$_3$N$_4$ membrane. We measure the thermal recovery time following photoexcitation by a 35-fs, 800-nm pump laser beam with a continuous-wave (cw) helium-neon probe and track the dynamics in real time. The microsecond resolution of this setup cannot resolve the initial femtosecond transient dynamics that result from pump excitation but instead focuses on the slow recovery dynamics of the system. The pump beam was focused to a spot size of $437 \times 341$ μm$^2$ (full width at half maximum), and the repetition rate was set to 50 Hz for conditions comparable to those in electron diffraction measurements [7,8]. The probe laser beam is split in two, with one beam passing through the VO$_2$ film, focused to be approximately 10 times smaller than the pump, and the second beam acting as an intensity reference. Both beams are detected on a balanced diode, and the reference intensity is adjusted to give zero voltage difference with the sample arm when there is no pump excitation of the sample. Thus, deviations from zero correspond to changes in the transmissivity of the VO$_2$ sample relative to the value at room temperature. These deviations are digitized with microsecond resolution by a fast oscilloscope.

Figure 1 shows the results obtained with the sample at atmospheric pressure [(a)–(d)] and with the sample in vacuum [(e)–(h)]. Both cases show qualitatively similar responses. At low fluences [Figs. 1(b) and 1(f)] below the phase-transition threshold, the transmission is reduced after photoexcitation, followed by an exponential recovery. This negative change, which increases in amplitude with fluence for values below the threshold, is consistent with a small temperature increase, as the transmissivity of $VO_2$ at 633 nm decreases when the temperature increases but remains below $T_c$ [19].

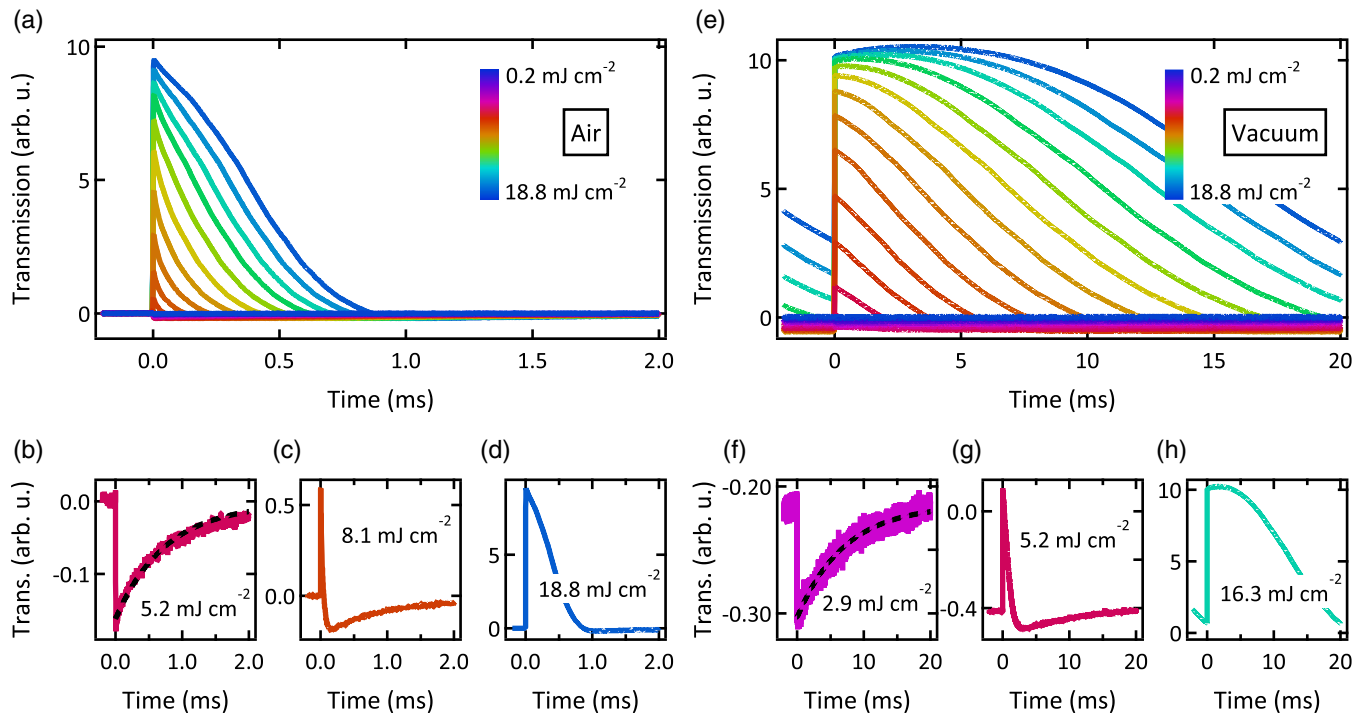

FIG. 1. Cooling dynamics of freestanding VO$_2$ films. (a) Optical response at 50 Hz, room temperature, and atmospheric pressure. Three fluence regimes can be distinguished: below (b) and above (c) the phase-transition threshold, and the saturation regime (d), where the optical properties no longer change with increasing fluence. Much slower relaxation dynamics can be observed when the sample is held in vacuum (e), which leads to lower values for the thresholds of the fluence regimes. Below the phase-transition threshold (f), the system shows a longer recovery time, and 5.2 mJ cm$^2$ is sufficient to induce the phase transition (g). The offset present at negative time delays is positive (h) if the sample remains partially metallic when the next pump pulse arrives.

For higher fluences [Figs. 1(c) and 1(g)], a sharp positive spike emerges after photoexcitation, which decays much more rapidly than the low-fluence signal. Yet, once decayed, the same slow negative signal is observed. This behavior implies that the threshold has been surpassed and the metallic phase has been partially induced in the system, as metallization results in an increased transmission of the probe light. As the fluence further increases, the magnitude of the positive change grows by over an order of magnitude and persists for longer times before eventually saturating [Figs. 1(d) and 1(h)]. At this point, the magnitude of the change no longer increases, consistent with the whole probed volume becoming metallic and remaining in this phase on the microsecond timescale.

While both the in-air and in-vacuum responses are qualitatively similar, there are key differences. Most striking is that the recovery time increases by an order of magnitude when the sample is in vacuum, rising from about 0.75 ms to about 7 ms. In vacuum, the sample no longer recovers between pulses, and heat accumulates, as evidenced by the significant offset from zero existing before the pump arrives, even for low-fluence excitation [compare the offsets in Figs. 1(a) and 1(f)]. This vacuum effect also allows us to rule out significant nonthermal contributions to the optical signal, such as long-lived excited carriers, which should be independent of the vacuum conditions.

These observations show that air acts as the dominant cooling pathway in freestanding films, and without it, a significant amount of heat accumulates in the sample. Although pump-probe experiments can still be performed, the initial state of the system now depends on the power deposited in the sample and thus the excitation fluence. This case can only be avoided by moving to sub-Hz repetition rates. As electron diffraction experiments have to be performed in vacuum and optical experiments are typically performed in air, differences in heating can lead to significant systematic errors.

To understand how this effect can influence our understanding of the phase-transition threshold, we plot the fluence dependence of the signal obtained at 10 μs with the sample in air (solid red circles) and in vacuum (open red circles) in Fig. 2(a). The threshold for the phase transition clearly shifts by almost 5 mJ cm$^{-2}$, which can be attributed solely to heat accumulation as all other experimental conditions remain the same. In vacuum, changing fluence now changes both the amount of energy provided by the laser in each pulse *and* the sample temperature. As hotter samples transform with a lower fluence [17], the threshold fluence shifts to lower values. This thermal buildup and establishment of a new quasiequilibrium temperature is plotted in real time in Fig. S1 of the Supplemental Material [20].

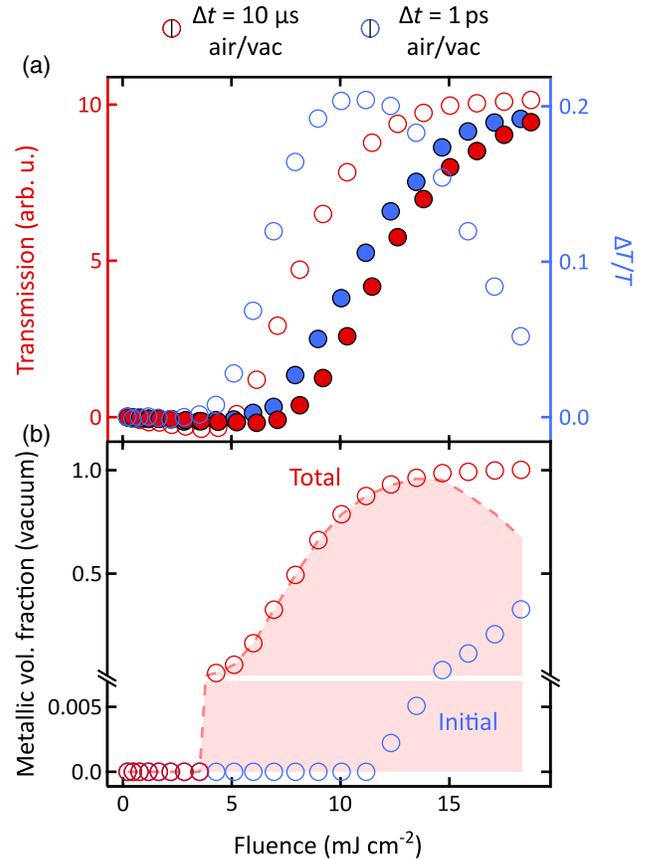

FIG. 2. (a) Fluence dependence transmission of the freestanding thin film of $VO_2$ in the microsecond (red) and femtosecond (blue) regimes, measured both at atmospheric pressure (filled) and in vacuum (empty). The shift in threshold fluence due to vacuum demonstrates that accumulated heat strongly modifies the perceived threshold fluence. (b) Volume fraction of the rutile R phase as a function of temperature for the sample in vacuum. Red open circles correspond to the total volume fraction generated at the sample 10-μs after excitation. The blue open circles correspond to the volume fraction that existed before the pump pulse arrived. The dashed line is the difference between these data, and it corresponds to the pump-induced volume fraction.

To ensure that the long-timescale dynamics measured here have the same effect on the processes that occur at femtosecond timescales, we switched the probe beam to a femtosecond optical probe at the same central wavelength (630 nm) and measured the transient signal in a conventional pump-probe experimental arrangement while keeping the pumping configuration identical. The fluence dependence of the data taken at a delay of 1 ps is also shown in Fig. 2(a). Despite the six orders of magnitude difference in timescales, the threshold values found are in remarkable agreement, and the same shift in the threshold value is observed between vacuum and air. The small reduction in threshold for the picosecond data relative to the cw data indicates there is a narrow fluence regime in which the metallic phase is generated but recovers within the 10-$\mu$s time resolution of the cw measurement and is therefore not observed at longer times.

When the sample is in vacuum, the picosecond fluence dependence peaks, before diminishing, which is not seen in the cw data. This case is due to differences in the measured quantities. In the microsecond data, the reference transmission is fixed eternally, while the femtosecond response is measured with a 100-Hz probe, with the pump pulse set at 50 Hz. The differential measurement is constructed from the difference between the transmitted intensities in phase (signal) and out of phase (reference) with respect to the pump pulse. If the sample makes a complete recovery, the out-of-phase reference and the external reference should match. However, as there is heat accumulation, the signal measured in the reference channel also changes with fluence. If the heating is sufficient to induce the metallic phase in the reference channel, the magnitude of the pump-induced change will be reduced.

In Fig. 2(b), we plot the fluence dependence of the rutile phase volume fraction, extracted from the microsecond optical data in vacuum, immediately prior to (initial, blue) and after (total, red) photoexcitation. This shows that at about 12 mJ cm$^{-2}$, the sample has been heated sufficiently so that some regions have already become metallic, and the system is now in a phase-separated state (see the Supplemental Material and Fig. S2 for more details [20]).

These results demonstrate that experimental conditions strongly influence the thermal recovery of the system. Heat accumulation can dramatically affect the measured phase-transition fluence threshold and even the starting phase fractions of the material. Although not shown here, we have confirmed that the recovery time and hence the measured threshold fluence change with spot size as expected for a thermal process [21]. The amount of heating depends on several experimental factors, specifically the laser repetition rate, spot-size, and environmental surroundings. All of these parameters often change between experiments, which makes accurately comparing fluence values challenging. For example, the claim of a monoclinic metallic state in Ref. [7] was based on comparing the thresholds obtained from a mid-IR probe experiment performed in air and an electron diffraction measurement in vacuum. In Ref. [8], the fluence dependence of the THz transmission was measured in air, while the electron diffraction measurements were performed in vacuum. Furthermore, spot sizes and repetition rates were not the same in the different experiments.

As demonstrated here, such measurements and their comparisons are likely compromised by a strong heat accumulation contribution that will depend on the experimental technique. Therefore, differences in thresholds, which represent key evidence for the presence of a transient monoclinic metallic phase, could easily have arisen from the inadvertent heating of the sample rather than from two distinct phase-transition mechanisms. Thus, in order to make meaningful comparisons of threshold fluences for different degrees of freedom, they must be measured under identical experimental conditions and ideally simultaneously, to prevent any systematic errors.

In the following sections, we present results from time-resolved XAS spectroscopy on freestanding films, which has the advantage of being sensitive to both electronic and lattice properties in the same technique. Although heating effects are certainly present in these samples, the ability to simultaneously extract both structural and electronic information enables us to show that both transitions occur at the same threshold fluence, despite the elevated temperature. In addition, we perform broadband optical probing above and below the band gap on the same sample in air, with parameters that result in insignificant heating. This method again shows a single threshold fluence, indicating a single photoinduced phase transition.

## III. ELECTRONIC AND STRUCTURAL CHANGES AT THRESHOLD

### A. XAS spectroscopy

X-ray absorption spectroscopy is an ideal tool for studying both electronic and structural changes in materials. It probes transitions from localized core states into unoccupied levels of the conduction band and is sensitive to changes in the crystal field and shifts in the Fermi level. The oxygen $K$-edge of VO$_2$ has been thoroughly studied, and changes in this spectral region through the phase transition are well understood. These XAS features have been used to highlight the role of electronic correlations [22,23], to identify the presence of other insulating phases [24], and to provide contrast in images of phase separation with high spatial resolution [25]. Dynamics in the soft x-ray region have been previously measured in transmission on picosecond [26] and femtosecond timescales [27] in freestanding samples. However, these measurements had low-energy resolution ($\sim$4 eV), and the high repetition rate and power of the pump laser employed would have undoubtedly heated the system into the metallic phase. As a result, true measurements of the soft x-ray dynamics of the insulator-metal phase transition in VO$_2$ have yet to be performed.

The static oxygen $K$-edge absorption spectra of our freestanding samples, measured in transmission with 0.1-eV spectral resolution, are shown in Fig. 3(a) for the insulating (300 K) and metallic (360 K) phases. A strong covalent hybridization occurs between the oxygen 2p and vanadium 3d levels of VO$_2$ [28], which gives rise to the main absorption features observed at the oxygen $K$-edge. The assignment of these features by Abbate *et al.* in 1991 [29], based on the model initially proposed by

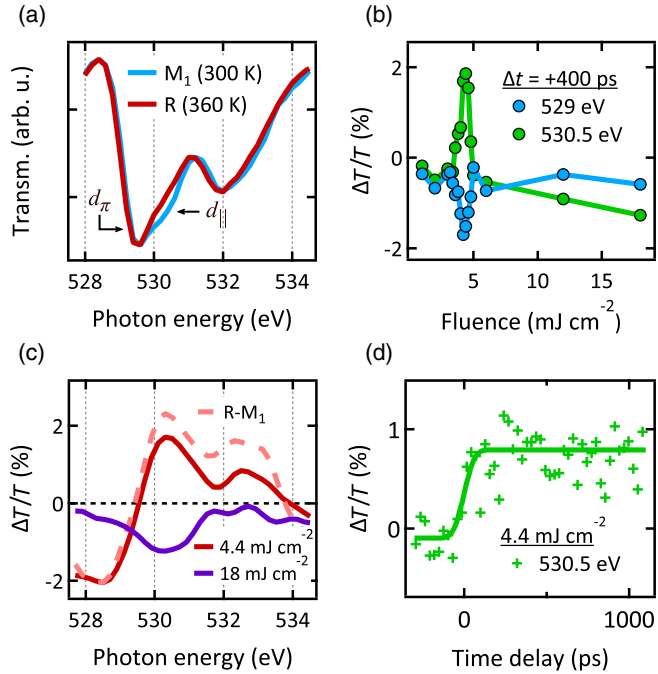

FIG. 3. Simultaneous measurements of the structural and electronic degrees of freedom with time-resolved x-ray absorption spectroscopy. (a) Static, high-resolution x-ray spectra measured at temperatures below and above $T_c$. The absorption peaks observed correspond to the $\pi^*$, $d_\parallel$, and $\sigma^*$ bands. (b) Fluence dependence of the $\pi^*$ (529 eV, blue) and $d_\parallel$ (530.5 eV, green) states. The insulator-metal threshold is found at 3 mJ cm$^2$, and a maximum change is observed at 4.4 mJ cm$^2$. High fluences result in a decrease in signal as the material is heated until it is already in the metallic phase upon excitation. At the largest fluences, a new XAS signal is observed, resulting from dynamics of the metallic phase. (c) Comparison of transient and thermal XAS spectra at the oxygen $K$-edge. The differential transient spectrum (solid red line) measured at a delay of 400 ps shows an excellent agreement with that obtained from thermal changes (dashed line). The metallic transient response (purple) has a completely different spectrum. (d) The temporal dynamics of the $d_\parallel$ (530.5 eV, green) states show changes rising within the 70-ps experimental resolution and show no subsequent dynamics. Note that this measurement was performed on a different region of the sample, which showed a smaller signal than (a)–(c).

Goodenough [30], has remained qualitatively unchanged, even after subsequent developments in the theoretical understanding of VO$_2$ [15,22,23]. For our purposes, the main features of interest are the $\pi^*$ and $d_\parallel$ states, present at 529 eV and 530.5 eV, respectively. The $\pi^*$ band is present in both insulating and metallic phases, and it shows a small redshift on transforming from the insulating to the metallic phase, as lower-energy transitions into states that were previously gapped out become allowed [22,28,31]. The $d_\parallel$ states are only present in the insulating phase, and they result from the structural dimerization of the vanadium ions along the rutile $c$ axis. The sensitivity of the $\pi^*$ state to the Fermi level means it can be used to monitor changes in the electronic properties of the material, while the $d_\parallel$ state probes the structural transformation [22,23,32]. In the dynamic regime, if a transient monoclinic metallic state emerges, we would expect to observe a change in the $\pi^*$ states for a pump fluence below that required to induce a change in the $d_\parallel$ states. As these features can be measured in the same experimental setup and under the same conditions, we can track both degrees of freedom without ambiguity.

We measured the transient XAS signal at the FemtoSpeX facility (UE56/1 ZPM) at BESSY II [33,34] with 1-eV spectral resolution and 70-ps time resolution. The static XAS measured at the beamline are shown in Fig. S5 of the Supplemental Material [20] together with a comparison to the high resolution measurements. The energy resolution of these transient data is a factor of 4 improvement on previous measurements [26,27]. While the time resolution is insufficient to capture the initial electronic [35] or lattice dynamics [36], it is still sufficient to observe the claimed monoclinic metallic state with a proposed lifetime greater than 600 ps [8]. To minimize heating, we operate at the minimum pump repetition rate of the facility, 600 Hz. However, under these conditions, heat accumulation is still significant, so additional cooling with liquid nitrogen was used to keep the base temperature of the sample below $T_c$ for most pump fluences. Optical characterization of the sample under these experimental parameters (displayed in the Supplemental Material, Fig. S4 [20]) showed that the system remains in the monoclinic phase for pump fluences below approximately 5 mJ cm$^{-2}$.

Figure 3(b) shows the pump fluence dependence of the $\pi^*$ and $d_\parallel$ states, probed at 529 and 530.5 eV, respectively, at a delay of 400 ps after pump excitation, within the lifetime of the proposed monoclinic metallic phase. No discernible changes in either feature are observed until the excitation reaches 3 mJ cm$^{-2}$, at which point the transmission at 529 eV begins to rapidly decrease while, at the same time, the transmission at 530.5 eV increases. Both signals grow in amplitude up to a fluence of 4.4 mJ cm$^{-2}$, where their maxima are reached. For higher fluences, both signals decrease until no transient change is observed at 5 mJ cm$^{-2}$, in a similar way to the picosecond optical data displayed in Fig. 2(a). Again, the suppression of the pump-probe signal results from the fact that the metallic phase has been thermally induced for high pump fluences.

In Fig. 4, we extract the volume fraction of the rutile metallic phase based on the responses at 529 eV and 530.5 eV independently and assuming that only the rutile phase is generated, in a similar approach to Ref. [13]. In Fig. 4(a), we show the phase fraction that exists immediately after photoexcitation, which includes both the transiently generated signal and the thermal signal. Figure 4(b) shows the long-lived thermal phase fraction only. The shaded area corresponds to the fluence range in which a transient pump probe signal is observed, i.e., the region in which some M$_1$ phase is still present in the initial state and can be converted into the rutile phase upon photoexcitation.

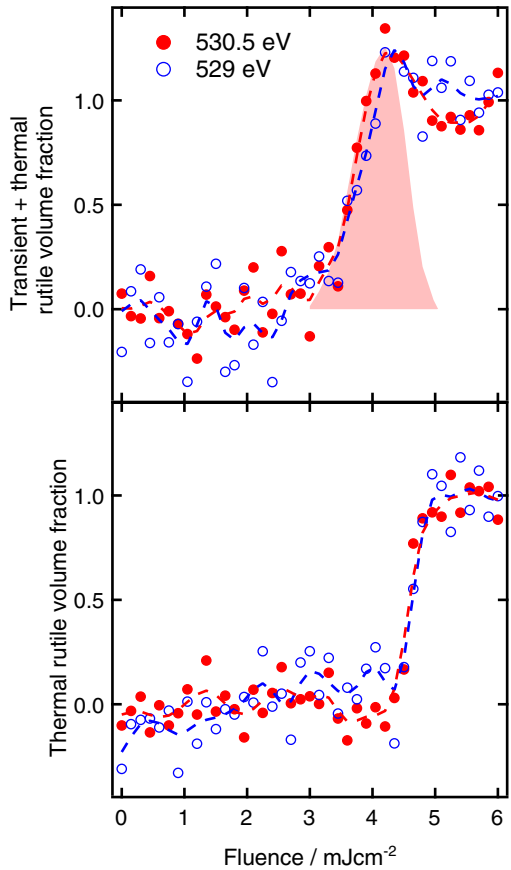

FIG. 4. Fluence dependence of the rutile phase volume fraction obtained from the XAS response at 530.5 eV (filled red circles) and 529 eV (blue open circles). (a) Maximum rutile phase volume fraction deduced from the XAS spectra measured 400 ps after photoexcitation. This fraction includes both the transiently induced change and the quasistatic thermal change. (b) Volume fraction obtained 4 ms after excitation, which shows the quasistatic volume fraction of the rutile phase. The dashed lines show the data smoothed. The shaded area corresponds to the region in which a transient phase transition is observed.

If a third phase were generated with a different XAS spectrum to either the M$_1$ or R phases, we would expect to see discrepancies between the volume fractions obtained from the 530.5-eV and 529-eV edges. For instance, if a transient metallic monoclinic phase was formed simultaneously with a transient rutile phase, we would expect the 529-eV signal to show a larger metallic volume fraction than the 530.5-eV signal. However, both estimates are in good agreement, leaving little room for a transient metallic monoclinic phase to coexist. The complete transient XAS spectrum at the fluence exhibiting the largest pump-induced change in volume fraction are shown in Fig. 3(c) (red trace). This spectrum is in excellent agreement with the thermal transition (dashed line), as expected from the extracted volume fractions. In addition, Fig. 3(d) shows that the dynamics are unchanging on the timescale measured, indicating the system has already thermalized on the timescales measured here. Therefore, our results are consistent with a single phase transition from the insulating M$_1$ phase to the rutile metallic phase.

Interestingly, for even higher pump fluences, a new transient change in the XAS signal emerges, which is shown in Fig. 3(c) (purple trace). This signal corresponds to the response of the photoexcited metallic phase and is not a new phase, as no threshold is observed. We believe this case would correspond to the states previously measured in Refs. [26,27].

### B. Optical signatures of the threshold

To further examine the nature of the transiently generated phase, we also investigate the optical response. The insulator-metal phase transition in VO$_2$ has been studied extensively with ultrafast optics and, although optical techniques lack the direct specificity of other approaches, spectroscopy can be harnessed to achieve greater sensitivity to structural or electronic properties. Two key energies to consider are the band gap of the M$_1$ phase, at 0.6 eV (2 μm) [37], and the plasma frequency of the metallic phase at 1 eV, or 1.25 μm [38]. Probing at, or below, this spectral region provides a strong sensitivity to changes in the conductive response of free electrons.

In contrast, photons with energies much higher than the band gap or plasma frequency are less sensitive to the free electrons. The region around 2.5-eV (500-nm) probes transitions between occupied oxygen states and unoccupied states of predominantly vanadium character near the Fermi level [37]. These transitions are sensitive to the structure, as changes in the crystal field modify the position and overlap of the bands. Previous measurements have shown that when the M$_1$ phase is excited below the phase-transition threshold, coherent phonons strongly modulate the transient reflectivity in this region. Furthermore, the coherent phonon signal has been used to show that the phonons are lost at the same fluence in which the reflectivity change indicates the phase-transition threshold [19,39].

We exploit this wavelength-dependent sensitivity to observe how the electronic and structural properties change after photoexcitation. If a monoclinic metallic phase exists with a threshold fluence lower than the fluence required to drive the system into the rutile metallic phase, we expect to observe a lower threshold in the optical response when probing with longer wavelengths. However, each probe is unlikely to be exclusively sensitive to either the metallic transition or the structural transitions. Thus, we also expect to find some evidence for two thresholds at all probe wavelengths.

Experiments were performed on the same sample as discussed so far. However, experiments were performed in air, so heating was not an issue and 100-Hz repetition rates could be used to improve the signal-to-noise ratio. Probe wavelengths spanning 680 nm–5 μm were generated by an optical parametric amplifier and further nonlinear optics. The change in transmission at a time delay of 1 ps was recorded as a function of pump fluence for each probe wavelength and is shown in Fig. 5(a).

As with the XAS measurements, we observe a single fluence threshold, which is the same for all probe wavelengths. In order to further demonstrate that the visible region is sensitive to structural changes, we show the coherent phonon amplitude and reflectivity transient obtained from a single crystal of $VO_2$ in Fig. 5(b). We see that the coherent phonon is suppressed at the same threshold as the background optical data, in agreement with previous measurements [19,39] (see Fig. S3 of the Supplemental Material [20] for the corresponding time traces). When taken together with the XAS data, our results show that there is a single threshold fluence and that this fluence corresponds to both the electronic and structural transitions.

## IV. ULTRAFAST THERMAL ORIGIN OF THE PHASE TRANSITION

Our measurements show that a single fluence threshold exists in $VO_2$. Furthermore, the observed threshold value can be dramatically altered by heat. Yet, although it is clear that the initial temperature of the system prior to laser excitation plays a role in the photoinduced transition, the exact nature of the threshold fluence has yet to be clarified. In particular, there is still a debate over whether the threshold fluence corresponds to the thermal energy required to heat the system above the transition temperature [16,17,40] or if the ultrafast transition proceeds along a different nonthermal path [41–43].

For a thermal transition in a thick single crystal, the following relation is expected between the threshold fluence, $F_{TH}$, and the thermal energy required to drive the transition, $E_T$, assuming optical absorption is in the linear regime,

$$F_{TH} = \frac{E_T d}{(1-R)}, \qquad (2)$$

where $R$ is the reflectivity of the sample and $d$ is the optical penetration depth. The thermal energy to drive the phase transition can be obtained by integrating the heat capacity from the initial temperature up to the phase-transition temperature.

There are various challenges in evaluating this simple-looking relation accurately. Optical properties can change with strain and thin film quality [37,44]. Strain can also shift $T_c$ and, as the heat capacity diverges at the phase transition, it may be questionable to use values in the literature to calculate $E_T$ in this region. In an attempt to circumvent these issues, we perform measurements on single crystals of $VO_2$ that have been carefully characterized optically (see Fig. S7 of the Supplemental Material [20]). Instead of calculating the energy needed to bring the sample from room temperature up to the transition temperature, we calculate the energy to raise the sample from 78 K to room temperature and compare that value to the difference in fluence threshold required to trigger the phase transition at both temperatures. In this way, we avoid the rapidly diverging region of the heat capacity near $T_c$.

Figure 6 shows the fluence dependence of the insulator-metal phase transition in the single crystal for two initial temperatures, 78 K and 295 K. We find that the temperature change results in a threshold shift by 3.4 mJ cm$^{-2}$. To calculate the thermal energy to raise the sample from 78 K to 295 K, we integrate the heat capacity, reported in Ref. [45], over the desired temperature range to get $E_T = 432 \text{ J cm}^{-3}$. From the optical characterization of a similar sample, we obtain a penetration depth of $d = 62$ nm and a reflectivity of $R = 0.18$. If the phase transition were driven by thermal energy, we would need to provide an additional fluence of 3.3 mJ cm$^{-2}$ to the system to trigger the insulator-metal phase transition for a sample initially at 78 K compared to 295 K. This case is in excellent agreement with the shift of 3.4 mJ cm$^{-2}$ that we observe, indicating that the critical fluence corresponds to the energy required to drive the system over the thermal barrier.

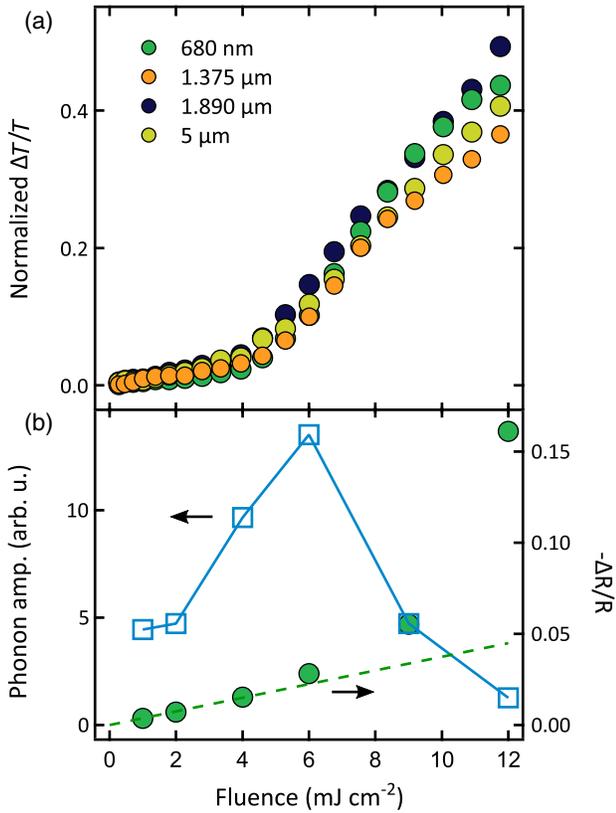

FIG. 5. (a) Normalized change in transmissivity of a freestanding $VO_2$ sample at 1-ps time delay measured at 100 Hz in air for different probe wavelengths. Below- and above-gap probing show the same threshold fluence, indicating that both the electronic and structural transitions occur at the same critical fluence. (b) Coherent phonon amplitude as a function of fluence measured at 690 nm from a single-crystal sample, compared with the reflectivity change measured at 1 ps, indicating the threshold measured in the visible spectral range corresponds to the structural threshold. The dashed green line is a linear fit to the low-fluence reflectivity.

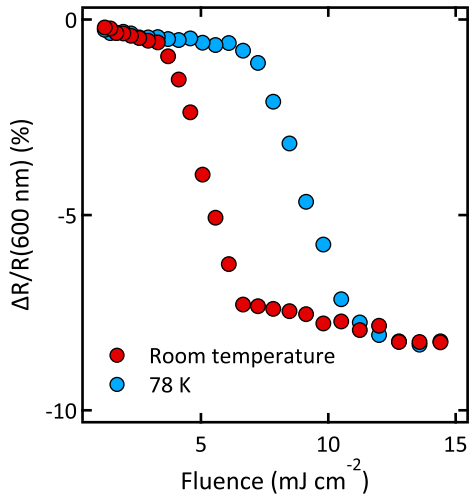

FIG. 6. Fluence dependence of the reflectivity change across the insulator-metal phase transition measured on a single crystal starting at an initial temperature of 78 K (blue circles) and 295 K (red circles), respectively. The difference in threshold can be explained by the thermal energy difference required to heat the sample from 78 K to 295 K.

## V. DISCUSSION

In this paper, we have reported measurements of the ultrafast phase transition in VO$_2$ probed at optical and soft x-ray wavelengths over a range of timescales. We have argued that spectroscopy can be used to tune the probe to be more sensitive to either the structural or electronic degrees of freedom within the same experimental technique and that, when this is done, both the structure and electronic degrees of freedom are found to change at the same critical fluence.

We have also shown that the threshold fluence to drive the ultrafast transition corresponds to the thermal energy required to drive the M$_1$-to-R phase transition. This result suggests that the ultrafast phase transition is dictated by the thermodynamic properties of the equilibrium. As strain can shift the phase-transition temperature, samples with different substrates should show different threshold fluences. Furthermore, we have also shown that heat accumulation can occur in freestanding samples, which causes significant changes in the observed threshold fluence in a nontrivial way. This case makes it challenging to compare thresholds between different samples or between the same sample measured in different experimental situations. Our results suggest that the difference in the absolute value of the threshold is related to differences in the samples and heating conditions, rather than the presence of a new phase.

Freestanding thin-film samples of VO$_2$, which are required for many ultrafast measurements, are particularly affected by cumulative heating. As a result, the initial sample temperature depends on the pump power used to excite the system. This not only modifies the fluence required to drive the phase transition but can also change the initial phase of the material. Thin films are known to undergo phase separation as a function of temperature [25,46–48]; thus, heating will drive the material away from a purely M$_1$ initial state. We have shown that there is already a significant volume fraction of a preexisting R phase for high-fluence excitation, but other thermodynamic phases such as M$_2$ or T [6] could also be present in the sample below our detectable limit. Such phase coexistence will influence the dynamics of the light-induced phase transition. For instance, when the initial state is purely M$_1$, the metallic phase must nucleate in the sample. On the contrary, if the phase coexistence is present in the sample, the preexisting metallic domains may grow, and these processes will have different dynamics [49]. This case may explain why the (200) peak is slow when observed with electron diffraction on freestanding films [7,8] but fast when observed with x-ray diffraction on single crystals [36]. Therefore, time-resolved nanoscale imaging techniques will be key to understanding how these different processes compete [50].

The thermal nature of the ultrafast phase transition raises questions about how to understand and ultimately control the process. Theoretical calculations suggest that it is possible, in principle, to close the band gap through the number of photoexcited holes alone, without the need for a structural change [9]. The approximately 0.6-eV band gap of VO$_2$ corresponds to a temperature of 6900 K, which suggests that the change in the number of thermally excited electrons should be negligible when compared to the number of electrons photoexcited at threshold ($\sim 10^{26}$ m$^3$ for 0.1 photoexcited electrons per vanadium). Therefore, a screening-driven band-gap collapse should be independent of the initial temperature, in stark contrast to what we observe. Moreover, for photon energies close to the band gap, lower-frequency light contains more photons per unit of energy than higher-frequency light. Thus, for hole-number-driven phase transitions, low-frequency light should be more energy efficient and require lower pulse energies. Yet, earlier work on the pump-wavelength dependence of the threshold indicated that the important quantity was the total absorbed energy [51], as expected for a thermal process.

While the ultrafast phase transition appears strongly connected to the thermal transition, it is still remarkable that it can occur so quickly. Two-thirds of the latent heat of the equilibrium transition can be attributed to phonon entropy [52]. If the photoinduced transition in $VO_2$ is thermal-like, then it points to a rapid, sub-100-fs transfer of energy from the excited electronic state to the lattice. Although this process is usually considered to be too slow to explain the rapid timescales observed in $VO_2$ [9], it is consistent with recent diffuse x-ray scattering measurements, which showed that phonons with a broad range of phonon wave vectors are excited at the phase-transition threshold [36]. This consistency points to the fact that the electron-phonon scattering rate may be higher due to a large lattice anharmonicity and phonon density of states in the excited system [52]. Fast electronic relaxation and energy transfer to phonons may explain why the theoretically suggested monoclinic metal has remained elusive, as the excited electrons simply do not remain in excited orbitals long enough for the screening effects to be observed. This work then challenges theory to accurately include realistic coupling to the phonon thermal bath, a degree of freedom that has been neglected to date, in order to make further progress in understanding light-induced phase transitions and whether this process is limited to $VO_2$ or applicable to a range of nonequilibrium phase transitions in other correlated materials.

## VI. CONCLUSION

We have performed an extensive range of optical and x-ray measurements on thin freestanding films and bulk single crystal samples of $VO_2$. We show that thin freestanding films are susceptible to average heating effects, which depend strongly on the experimental parameters. This case results in a direct dependence of the system base temperature on the pump fluence and makes comparing different experimental techniques difficult. When these effects are taken into account, or avoided, we observe a single threshold fluence that corresponds to both the structural and electronic insulator-metal phase transition. The observation of a single fluence threshold for both the structural and electronic transition brings the ultrafast phase transition in $VO_2$ into agreement with static measurements, which have ruled out the existence of a decoupled thermal phase transition [15]. While freestanding thin films are particularly susceptible to heating due to limited options for heat dissipation, the relaxation time in some single-crystal manganites can be even longer if the system is close to a glass transition [53]. Thus, similar effects may be present in a range of materials, and care is needed when comparing results obtained under different experimental conditions.

We have shown that the energy required to drive the photoinduced phase transition in $VO_2$, transiently, is comparable to the energy required to drive the system thermally. To date, most theoretical work on the light-induced phase transition has focused on the role of nonthermal electronic states and electronic correlations. However, it appears that a better theoretical understanding of how the initial thermal state of the lattice, and how it couples to excited electrons, will be required to make further progress in understanding the properties of $VO_2$ and potentially other quantum materials.

Finally, we note that $VO_2$, especially in thin films, can be inhomogeneous, with multiple insulating and metallic phases coexisting in equilibrium [25,46–48]. Therefore, a complete understanding of the phase transition will ultimately need to include how these phases compete dynamically in real space during the light-induced transition [50].

## ACKNOWLEDGMENTS

We thank Christian Schüßler-Langeheine, Torsten Kachel, and Niko Pontius for help during the beam time. This project has received funding from the European Research Council (ERC) under the European Union's Horizon 2020 Research and Innovation Programme (Grant Agreement No. 758461), by the Ministry of Science, Innovation and Universities (MCIU), State Research Agency (AEI) and European Regional Development Fund (FEDER) PGC2018-097027-B-I00, the "Severo Ochoa" program for Centres of Excellence in R&D (SEV-2015-0522), 1. Fundació Mir-Puig, Fundació Privada Cellex, and CERCA Programme/ Generalitat de Catalunya. D. S. acknowledges the Helmholtz Association for funding via the Helmholtz Postdoctoral Program PD-142. K. H. was supported by a grant from the National Science Foundation (Grant No. EECS-1509740).